\documentclass[11pt]{article}
\pdfoutput=1

\usepackage{jheppub} 
\usepackage[T1]{fontenc} 

\usepackage{tikz}
\usepackage{pbox}

\usepackage{amsmath}
\usepackage{booktabs}%
\usepackage{multirow}

\allowdisplaybreaks

\usepackage{graphicx}

\usepackage{caption}

\usepackage{subcaption}

\usepackage{float}

\makeatletter
\def\@fpheader{\relax}
\makeatother

\usepackage{color}
\usepackage{amsmath}
\usepackage{graphicx}
\usepackage{amssymb}
\usepackage{esint}

\def\L{\Lambda}
\def\S{\Sigma}

\def\nn{\nonumber}
\def\p{\partial}
\def\ls{\left[}
\def\rs{\right]}
\def\lc{\left\{}
\def\rc{\right\}}

\newcommand{\bi}{\begin{itemize}}
\newcommand{\ei}{\end{itemize}}

\def\nn{\nonumber}
\def\p{\partial}
\def\ls{\left[}
\def\rs{\right]}
\def\lc{\left\{}
\def\rc{\right\}}
\newcommand{\be}{\begin{eqnarray}}
\newcommand{\ee}{\end{eqnarray}}

\subheader{\ } 
\title{ 
Non-minimal 
scalar 
multiplets, 
supersymmetry breaking and dualities 
}

\author{Fotis Farakos}
\emailAdd{fotisf@mail.muni.cz}

\author{, Ond{\v{r}}ej Hul\'{i}k}
\emailAdd{ondra.hulik@mail.muni.cz}

\author{, Pavel Ko\v{c}\'{i}}
\emailAdd{pavelkoci@mail.muni.cz}

\author{\! and \!\!\! }

\author{Rikard von Unge}
\emailAdd{unge@physics.muni.cz}

\affiliation{ Institute for Theoretical Physics, Masaryk University, \\  611 37 Brno, Czech Republic}

\abstract{
We study supersymmetry breaking in theories with non-minimal multiplets (such as the 
complex linear or CNM multiplets), 
by using superspace higher derivative terms which give rise to new 
supersymmetry breaking vacuum solutions on top of the standard supersymmetric vacuum. 
We illustrate the decoupling of the additional massive sectors 
inside the complex linear and the CNM multiplets and show that  
only the Goldstino sector is left in the low energy limit.
We also discuss the duality between non-minimal scalar multiplets and chiral multiplets in the presence of superspace higher derivatives.  
{}From the superspace Noether procedure we calculate the supercurrents,
and we show that in the supersymmetry breaking vacuum the chiral superfield $X$ which enters 
the Ferrara-Zumino supercurrent conservation equation  
does indeed flow in the IR to the chiral constrained Goldstino superfield. 
We also provide a description of the Goldstino sector in terms of the Samuel-Wess superfield for the supersymmetry breaking mechanism at hand.
}

\begin{document} 
\maketitle
\flushbottom

\section{Introduction}

If supersymmetry \cite{Gates:1983nr} is realized in nature, 
it has to be spontaneously broken. 
It is  common practice to identify the supersymmetry breaking sector with some hidden sector, 
and its main impact in particle physics is solely the breaking of supersymmetry \cite{Martin:1997ns}. 
Therefore, 
the study of the various supersymmetry breaking mechanisms and the patterns they give 
for the breaking in the low energy, 
would in principle serve as a way to distinguish between the various possibilities. 
In this work we will study the  non-minimal superfields 
\cite{Gates:1980az,Deo:1985ix,Gates:1995fx,Gates:1996cq,Grisaru:1997hf,
Penati:1997pm,Penati:1997fz,TartaglinoMazzucchelli:2004vt,GonzalezRey:1997xp,GonzalezRey:1997qh,
Kuzenko:2011ti,Farakos:2013zsa,Farakos:2014iwa}, 
in 4D, $N=1$, 
as candidates for the supersymmetry breaking hidden sector.

Supersymmetry breaking by a pure  complex linear superfield contribution 
has only recently shown to be  possible \cite{Farakos:2013zsa,Farakos:2014iwa}\footnote{However in \cite{Kuzenko:2011ti} a different supersymmetry breaking mechanism using a modified complex linear superfield was studied.}. 
Even though a superpotential can not be used to deform the auxiliary field potential and  break supersymmetry  
it has been found that instead one may use superspace higher derivative terms to achieve this.  
In particular, a model which will do this is given by (in the conventions of \cite{Gates:1983nr}) 
\be
{\cal L} = -\int d^4 \theta \ \bar \Sigma \Sigma 
+ \frac{1}{8 f^2}  \int d^4 \theta  D^\alpha \Sigma D_\alpha \Sigma 
\bar D^{\dot{\beta}} \bar \Sigma  \bar D_{\dot{\beta}} \bar \Sigma .
\ee
The mechanism relies on the existence of several solutions to the auxiliary field equations 
which leads to multiple vacua with different properties.  
Among these vacua, 
there is the standard supersymmetric solution ($\langle D^2 \S | \rangle = 0$)
in which the physics is the same as in the free theory, 
but there also exist vacua which break supersymmetry ($\langle D^2 \S | \rangle \ne 0$).  
In this work we will further investigate this mechanism 
both for the complex linear superfield but also for the chiral non-minimal (CNM) \cite{Deo:1985ix} multiplet, 
which contains both a complex linear and a chiral superfield, 
where the complex linear constraint is modified using the chiral field.
The main advantage of the CNM multiplet is that the complex linear superfield can naturally
be given a mass.

A characteristic property of the supersymmetry breaking mechanism discussed in this paper is 
that the massless fermionic excitation generically associated with global supersymmetry breaking,
the Goldstino, is identified with a fermion which in the free theory is auxiliary. 
In the supersymmetry breaking vacuum it acquires a kinetic term and becomes propagating. 
This means that the superspace higher derivative term induces supersymmetry breaking 
while introducing additional propagating modes. 
Similar properties of supersymmetric theories, 
not related to supersymmetry breaking, 
have been found in a supergravity setup \cite{Cecotti:1987qe,Cecotti:1987sa,Farakos:2015hfa}. 
In a supersymmetric setting the Goldstino can be nonlinearly embedded in a
chiral superfield $X_{NL}$ \cite{Rocek:1978nb,Lindstrom:1979kq}.  
This superfield satisfies the constraints 
\be
\label{chG1} 
X_{NL}^2 &=& 0 
\\
\label{chG2}
\bar{X}_{NL} D^2 X_{NL} &=& f \bar{X}_{NL} 
\ee
which  remove the  scalar partner of the Goldstino from the spectrum 
and fix the vev of the auxiliary field to a non-vanishing value $f$. 
The  constraint \eqref{chG2} can be  implemented from the equation 
\be
\label{chG3}
D^2 X_{NL} = f  + \cdots 
\ee 
which will also yield an equation of motion for the Goldstino.

It is well known that  there exists a duality between models of complex linear superfields and chiral superfields.
The duality is robust in the sense that it does not rely on the existence of special properties of the model, such as
isometries in the case of sigma models. 
In fact, one might be tempted to conclude that the duality can always be
performed in any model built with complex linear superfields. 
However, the theories studied in this paper show that 
in the supersymmetry breaking vacuum the complex linear model has more degrees of freedom 
than what can be described by a single chiral superfield. 
The chiral-linear duality can still be performed in a setting where 
one perturbatively solves the equations of motion of the parent theory around the appropriate background. 
In this procedure the additional degrees of freedom, even though they are dynamical, are contained  in the background. We also discuss the appropriate Lagrangian description for these new degrees of freedom.

After describing the generic properties of the models, 
and finding the supersymmetry breaking vacua, 
we study their low energy limits. 
{}From the superspace Noether procedure \cite{Magro:2001aj}, 
we identify the $X$ superfield which enters the 
supercurrent  equation \cite{Ferrara:1974pz} 
\be
\bar D^{\dot \alpha} {\cal J}_{\alpha \dot \alpha} = D_\alpha X 
\ee
and we show that in the IR it flows to $X_{NL}$ \cite{Rocek:1978nb,Lindstrom:1979kq,Casalbuoni:1988xh} 
as has been advocated in \cite{Komargodski:2009rz}. 
More precisely, we calculate the Ferrara-Zumino (FZ) supercurrent multiplet, 
for both the complex linear model and the CNM,  
and we find that in the low energy limit
\be
X \rightarrow \frac13 f X_\text{NL} . 
\ee 
Since these theories have an exact R-symmetry, we also calculate 
the ${\cal R}$-multiplet \cite{Komargodski:2010rb,Kuzenko:2010am,Kuzenko:2010ni}. 
The existence of both the FZ-multiplet and also of the ${\cal R}$-multiplet,  
provides  evidence for the possibility of consistently coupling these models to 
the old-minimal and the new-minimal supergravity.

A different way of embedding the Goldstino in a superfield was invented in \cite{Ivanov:1978mx,Ivanov:1982bpa,Samuel:1982uh}. This procedure gives a realization of the Goldstino in terms of  a constrained spinorial superfield $\Lambda_\alpha$ where the constraints were explicitly given by Samuel and Wess in \cite{Samuel:1982uh}.
Already in \cite{Samuel:1982uh}, it was shown that the nonlinear embedding of the Goldstino into the chiral superfield $X_{NL}$
discussed above, can be realized using the Samuel-Wess superfield as
\be
X_{NL} \propto \bar{D}^2(\Lambda^2\bar{\Lambda}^2) . 
\ee
In this paper we argue that universally, for all models that break supersymmetry with a superspace higher derivative term
involving complex linear superfields, the Goldstino can be embedded in the complex linear superfield using the
SW-superfield as
\be
\S_{\Lambda} = \bar D^{\dot \alpha} \left( \bar \L_{\dot \alpha} \L^\alpha \L_\alpha \right)  .
\ee
This Goldstino superfield satisfies 
\be
\S_{\Lambda}^2 = 0  
\ee
and 
\be
\langle D^2 \S_{\Lambda} | \rangle \ne 0 
\ee
while it contains only the Goldstone fermion ($G_\alpha$), 
as a propagating mode in its lowest component $\L_\alpha | = G_\alpha$. 
We also discuss the superspace equations of motion implemented on $\L_\alpha$, 
from these models.

\section{Complex linear superfields and superspace higher derivatives}

In this section we study the supersymmetry breaking from the non-minimal superfields 
and comment on the duality to chiral superfields.

\subsection{CNM and supersymmetry breaking}

The CNM multiplet  \cite{Deo:1985ix} contains a complex linear superfield $\S$ as well as a chiral superfield $\Phi$ linked together
through the modified complex linear constraint
\be
\label{SMPHI}
\bar D^2 \S = m \Phi 
\ee
where $m$ is a mass scale. 
The component definitions are
\be
\Phi|=z \ , \   D_\alpha \Phi | = \rho_\alpha    \ , \ D^2 \Phi|= N  
\ee
and
\be
\nn
&&\S|=A \ , \ D^2 \S|= F \ , \ \bar D_{\dot \alpha} D_{\alpha} \S |= P_{\alpha \dot \alpha} , 
\\
\label{Scomponents}
&& \bar D_{\dot \alpha} \S|= \bar \psi_{\dot \alpha} \ , \ D_\alpha \S|= \lambda_\alpha \ , \ 
\frac12 D^\gamma \bar D_{\dot \alpha} D_{\gamma} \S |=\bar \chi_{\dot \alpha}   . 
\ee
In principle the component  fields $F$, $G$, $N$, $P_{\alpha \dot \alpha}$, $\chi_\alpha$ and $\lambda_\alpha$ 
are auxiliary and we integrate them out.  
The fields $\Phi$, $\S$  constrained by (\ref{SMPHI}) and the Lagrangian
\be
\label{Lfree}
{\cal L} = -\int d^4 \theta \ \bar \Sigma \Sigma 
+ \int d^4 \theta \ \bar \Phi \Phi
\ee
give the component Lagrangian (after we integrate out the auxiliary fields) 
\be
{\cal L} &=& \frac12 A \,  \p^{\alpha \dot \alpha} \p_{\alpha \dot \alpha} \bar A 
+\frac12 z \,  \p^{\alpha \dot \alpha} \p_{\alpha \dot \alpha} \bar z 
- m^2 z \bar z - m^2 A \bar A  
\\
\nn
&& -  i \psi_\alpha \p^{\alpha \dot \beta} \bar \psi_{\dot \beta} 
-  i \rho_\alpha \p^{\alpha \dot \beta} \bar \rho_{\dot \beta} 
-m \psi^\alpha \rho_\alpha -m \bar \psi^{\dot \alpha} \bar \rho_{\dot \alpha}  , 
\ee
and thus describes  a free massive theory.
Notice that the massive scalars $z$ and $A$ are accompanied by 
two massive Weyl spinors $\rho_\alpha$ and $\psi_\alpha$, 
which together constitute a massive Dirac spinor.

Now we turn to  supersymmetry breaking. 
The model we  study here is 
\be
\label{L1}
{\cal L} = -\int d^4 \theta \ \bar \Sigma \Sigma 
+ \int d^4 \theta \ \bar \Phi \Phi 
+ \frac{1}{8 f^2}  \int d^4 \theta  D^\alpha \Sigma D_\alpha \Sigma 
\bar D^{\dot{\beta}} \bar \Sigma  \bar D_{\dot{\beta}} \bar \Sigma  .  
\ee
To understand the vacuum structure we look at the bosonic sector of the theory, which  is 
\be
\nn
{\cal L}_\text{B} &=& \frac12 A \,  \p^{\alpha \dot \alpha} \p_{\alpha \dot \alpha} \bar A 
+\frac12 z \,  \p^{\alpha \dot \alpha} \p_{\alpha \dot \alpha} \bar z 
\\
&& - F \bar F 
+ P^{\alpha \dot \alpha}   \bar P_{\alpha \dot \alpha} 
- m^2 z \bar z - m A \bar N - m N \bar A + N \bar N  
\\
\nn
&& 
+  \frac{1}{2 f^2} F^2 \bar F^2 
+ \frac{1}{2 f^2} F \bar F P^{\alpha \dot \alpha}   \bar P_{\alpha \dot \alpha} 
+ \frac{1}{8 f^2} P^{\alpha \dot \alpha}    P_{\alpha \dot \alpha} \bar P^{\beta \dot \beta}   \bar P_{\beta \dot \beta} . 
\ee 
Since $N$, $F$ and $P_{\alpha \dot \alpha}$ are auxiliary fields, we integrate them out. 
By varying $N$ we get
\be
\bar N = m \bar A 
\ee
which then contributes to the total scalar potential 
\be
\label{scalarpotential}
V = m^2 z \bar z +m^2 A \bar A . 
\ee
{}From (\ref{scalarpotential}) we see that in the vacuum
\be
\langle z \rangle = 0 \ , \  \langle A \rangle = 0 
\ee
therefore $\langle N \rangle = 0$ . 
We now proceed to integrate out $P_{\alpha \dot \alpha}$. 
The variation with respect to  $P^{\alpha \dot \alpha}$ gives 
\be
\label{vectoreq}
\bar P_{\alpha \dot \alpha} + \frac{1}{2 f^2} F \bar F    \bar P_{\alpha \dot \alpha} 
+ \frac{1}{4 f^2}    P_{\alpha \dot \alpha} \bar P^{\beta \dot \beta}   \bar P_{\beta \dot \beta} =0  
\ee
which is  solved for 
\be
\label{vector1}
\langle P_{\alpha \dot \alpha} \rangle = 0  . 
\ee
There is also the solution 
$\langle P^{\beta \dot \beta}   \bar P_{\beta \dot \beta} \rangle = - 4 f^2 - 2 \langle F \bar F   \rangle $ 
where 
$P^{\beta \dot \beta} = \bar P^{\beta \dot \beta}$,  
which however we do not consider further in this paper. 
In the following we will always take the solution (\ref{vector1}). 
Finally, we want to integrate out $F$.

The variation with respect to $F$ gives
\be
\label{Feq}
- \bar F 
+  \frac{1}{ f^2} F \bar F^2  = 0 . 
\ee
It is easy to check that equation (\ref{Feq}) has two  solutions 
\begin{enumerate}

\item The standard supersymmetric solution with $\langle F \rangle=0$. 
Here supersymmety is not broken and $\langle V \rangle =0$ . 

\item The supersymmetry breaking solution with $\langle F \bar F \rangle= f^2$. 
Here supersymmetry is broken and $\langle V \rangle = \frac{f^2}{2}$ . 

\end{enumerate}
We have also included the vacuum energy of the theory, 
in the two vacua, 
such that the relation to supersymmetry breaking in evident.

The basic signal for supersymmetry breaking is the existence of a fermionic Goldstone mode.
{\em i.e.} 
the existence of a fermion which transforms with a shift under a supersymmetry transformation 
around the supersymmetry breaking vacuum. 
To understand the structure of the fermions we give the fermionic sector up to quadratic order 
\begin{align}
{\cal L}_\text{Quad.F} = - i \psi_\alpha \p^{\alpha \dot \beta} \bar \psi_{\dot \beta} 
&-  i \rho_\alpha \p^{\alpha \dot \beta} \bar \rho_{\dot \beta} 
+ \chi^\alpha \lambda_\alpha + \bar \chi^{\dot \alpha} \bar \lambda_{\dot \alpha} 
-m \psi^\alpha \rho_\alpha -m \bar \psi^{\dot \alpha} \bar \rho_{\dot \alpha} 
\\
\nn
+ \frac{1}{8 f^2} \Big{\{} &- 2 i (\p^{\alpha \dot \beta}  F) P_{\alpha \dot \beta} 
\bar \lambda^{\dot \gamma} \bar \lambda_{\dot \gamma} 
- 4 (i \p^{\alpha \dot \beta} \lambda_\beta 
- \frac{i}{2} \delta_{\beta}^{\alpha} \p^{\gamma \dot \beta} \lambda_\gamma 
- \delta^{\alpha}_{\beta} \bar \chi^{\dot \beta} ) 
P_{\alpha \dot \beta} \bar P^{\beta \dot \alpha} \bar \lambda_{\dot \alpha} 
\\
\nn
&- P^{\alpha \dot \beta} P_{\alpha \dot \beta} \bar \lambda^{\dot \alpha} 
(2 i \p_{\alpha \dot \alpha} \psi^\alpha + 2 m \bar \rho_{\dot \alpha} )
\\
\nn
&- \delta^\alpha_\beta F (2 m \delta_{\alpha}^{\beta} N -2 i \p_{\alpha \dot \beta} P^{\beta \dot \beta} 
- 2 \p_{\alpha \dot \beta} \p^{\beta \dot \beta} A )  \bar \lambda^{\dot \gamma} \bar \lambda_{\dot \gamma} 
\\
\nn
&- 2 \delta^\alpha_\beta F (2 m \rho_\alpha -2 i \p^{\gamma \dot \alpha} \bar \psi_{\dot \alpha} C_{\gamma \alpha}) 
\bar P^{\beta \dot \alpha} \bar \lambda_{\dot \alpha}   
\\
\nn
&- 2 \lambda^\alpha  (2 m \delta_{\alpha}^{\beta} N -2 i \p_{\alpha \dot \beta} P^{\beta \dot \beta} 
- 2 \p_{\alpha \dot \beta} \p^{\beta \dot \beta} A ) \bar P_{\beta \dot \alpha} \bar \lambda^{\dot \alpha}
\\
\nn
&- \lambda^\alpha (2 m \rho_\alpha -2 i \p^{\gamma \dot \alpha} \bar \psi_{\dot \alpha} C_{\gamma \alpha})  
\bar P^{\beta \dot \alpha} \bar P_{\beta \dot \alpha}
\\
\nn
&+ 4 (i \p^{\alpha \dot \beta} \lambda_\beta 
-\frac{i}{2} \delta_{\beta}^{\alpha} \p^{\gamma \dot \beta} \lambda_\gamma 
-\delta_{\beta}^{\alpha} \bar \chi^{\dot \beta}  ) \delta^\beta_\alpha F \bar \lambda^{\dot \alpha} 
C_{\dot \alpha \dot \beta} \bar F 
\\
\nn
&+4 (i \p^{\alpha \dot \beta} \lambda_\beta 
-\frac{i}{2} \delta_{\beta}^{\alpha} \p^{\gamma \dot \beta} \lambda_\gamma 
-\delta_{\beta}^{\alpha} \bar \chi^{\dot \beta}  ) \lambda_\alpha 
\bar P^{\beta \dot \alpha} C_{\dot \alpha \dot \beta} \bar F 
\\
\nn
&+ 4 P^{\alpha \dot \beta} C_{\alpha \beta} F \bar \lambda^{\dot \alpha} 
(\frac12 C_{\dot \alpha \dot \beta} \p^{\beta \dot \gamma} \bar \lambda_{\dot \gamma} 
+ C_{\dot \beta \dot \alpha} \chi^\beta ) 
\\
\nn
&- 4 P^{\alpha \dot \beta} \lambda_\alpha  \bar P^{\gamma \dot \alpha} 
(\frac12 C_{\dot \alpha \dot \beta} \p^{\beta \dot \gamma} \bar \lambda_{\dot \gamma} 
+ C_{\dot \beta \dot \alpha} \chi^\beta ) C_{\beta \gamma} 
\\
\nn
&- 2 P^{\alpha \dot \beta} \lambda_\alpha (-2 i \p^{\beta \dot \alpha} \psi_\beta +2 m \bar \rho^{\dot \alpha} ) 
C_{\dot \alpha \dot \beta} \bar F 
\\
\nn
&- 2 P^{\alpha \dot \beta} \lambda_\alpha \bar \lambda^{\dot \alpha} 
(2 i C_{\dot \alpha \dot \beta} \p^{\beta \dot \rho} \bar P_{\beta \dot \rho} 
+C_{\dot \beta \dot \alpha } \p^{\beta \dot \rho} \p_{\beta \dot \rho} \bar A 
- 2 m C_{\dot \beta \dot \alpha} \bar N )
\\
\nn
&- 4 \lambda^\alpha C_{\alpha \beta} F 
( - \chi^\beta + i \p^{\beta \dot \beta} \bar \lambda_{\dot \beta} )  \bar F 
+ 4 \lambda^\beta F \bar F ( - \chi^\gamma + i \p^{\gamma \dot \beta} \bar \lambda_{\dot \beta} ) C_{\gamma \beta} 
\\
\nn
&- 2 \lambda^\alpha \lambda_\alpha  \bar F 
(2 i \p^{\beta \dot \rho} \bar P_{\beta \dot \rho} 
- \p^{\beta \dot \rho} \bar \p_{\beta \dot \rho} \bar A + 2 m \bar N ) \Big{\}}. 
\end{align}
To find the propagating  modes in the two vacua, 
we write down the theory in the appropriate background and expand to quadratic order in the fields.

We start with the standard vacuum with $\langle F \rangle =0$. 
There we have the exact solution $F=0$ and $P_{\alpha \dot \alpha} =0$, 
which leads to 
\be
\label{intact}
{\cal L} &=& \frac12 A \,  \p^{\alpha \dot \alpha} \p_{\alpha \dot \alpha} \bar A 
+\frac12 z \,  \p^{\alpha \dot \alpha} \p_{\alpha \dot \alpha} \bar z 
- m^2 z \bar z - m^2 A \bar A  
\\
\nn
&& -  i \psi_\alpha \p^{\alpha \dot \beta} \bar \psi_{\dot \beta} 
-  i \rho_\alpha \p^{\alpha \dot \beta} \bar \rho_{\dot \beta} 
-m \psi^\alpha \rho_\alpha -m \bar \psi^{\dot \alpha} \bar \rho_{\dot \alpha} 
+ \chi^\alpha \lambda_\alpha + \bar \chi^{\dot \alpha} \bar \lambda_{\dot \alpha}  . 
\ee
Once we  integrate out the auxiliary fermions $\chi_\alpha$ and $\lambda_\alpha$, 
they will work as Lagrange multipliers for each other which will put them to zero, 
leaving behind two massive scalar multiplets, with Dirac mass for the fermions.  
In other words we recover the free theory we started with and with no trace of the higher dimension operator left. 
Note that this is an exact result, not an approximation. 
We will clarify  this  later using superspace methods.

In the supersymmetry breaking vacuum we have
\be
\langle F \bar F  \rangle = f^2 \ , \ \langle P_{\alpha \dot \alpha} \rangle=0 . 
\ee
The quadratic contributions  in this vacuum are  
\be
\label{218}
{\cal L}_\text{Quad.} &=& \frac12 A \,  \p^{\alpha \dot \alpha} \p_{\alpha \dot \alpha} \bar A 
+ \frac12 z \,  \p^{\alpha \dot \alpha} \p_{\alpha \dot \alpha} \bar z 
- m^2 z \bar z - m^2 A \bar A  
\\
\nn
&&-  i \psi_\alpha \p^{\alpha \dot \beta} \bar \psi_{\dot \beta} 
-  i \rho_\alpha \p^{\alpha \dot \beta} \bar \rho_{\dot \beta} 
-m \psi^\alpha \rho_\alpha -m \bar \psi^{\dot \alpha} \bar \rho_{\dot \alpha} 
\\
\nn
&&- \frac12 f^2 - i \lambda_\beta \p^{\beta \dot \beta} \bar \lambda_{\dot \beta} . 
\ee
{}From the Lagrangian (\ref{218}) we see that on top of the massive sector, 
there is a new fermionic mode in the last line (and we have also kept the positive vacuum energy manifest). 
In fact, the new mode is the previously auxiliary fermion which has acquired a kinetic term 
and taken the role of the Goldstino 
which transforms under a supersymmetry transformation in the supersymmetry breaking vacuum as 
\be
\delta \lambda_\alpha = f \,  \epsilon_\alpha + \cdots  
\ee 
The fact that there exists a Goldstino is dictated by supersymmetry breaking. 
%We will see this theory contains a Goldstino superfield on-shell. 

If we instead study the superspace formulation of the theory, we can derive the equations of motion 
from the Lagrangian (\ref{L1}) 
\be
\nn
{\cal L} &=& -\int d^4 \theta \ \bar \Sigma \Sigma 
+ \int d^4 \theta \ \bar \Phi \Phi 
+ \frac{1}{8 f^2}  \int d^4 \theta  D^\alpha \Sigma D_\alpha \Sigma 
\bar D^{\dot{\beta}} \bar \Sigma  \bar D_{\dot{\beta}} \bar \Sigma 
\\
&& + \int d^2 \theta  \ Y \left( \bar D^2 \S - m \Phi \right) + \int d^2 \bar \theta \ \bar Y \left( D^2 \bar \S - m \bar \Phi \right) 
\ee
where now $Y$ is a chiral superfield but $\S$ is unconstrained. 
By integrating out $Y$ we get (\ref{SMPHI}). 
If we on the other hand vary with respect to $\bar\S$ we get
\be
\label{eomS}
-  \S + \bar Y -  \frac{1}{4 f^2} \bar D^{\dot\alpha}  \left( \bar D_{\dot\alpha} \bar\S 
D^{\alpha} \S  D_{\alpha}\S \right) =0 . 
\ee
If we introduce a complex superfield $H$ satisfying
\be
\label{eomH}
  H +  \frac{1}{4 f^2} \bar D^{\dot\alpha}  \left(\bar D_{\dot\alpha} \bar H 
D^{\alpha} H  D_{ \alpha} H \right) =0 
\ee
and use the fact that $Y$ is chiral, we see that
\be
\label{sol1}
\S = \bar Y  + H 
\ee
solves the equation of motion (\ref{eomS}). With this redefinition we have separated the degrees of freedom from the
original complex linear field $\S$ into an antichiral field and the constrained field $H$. The equation for $H$ (\ref{eomH})
has several solutions. 
First, there is the trivial solution $H=0$ 
which corresponds to the supersymmetric vacuum. 
But there is also the solution 
where $\langle F \bar F \rangle \ne 0$ \cite{Farakos:2013zsa}, in which
\be
\label{HeqXNL}
H = X_{NL}
\ee
where  $X_{NL}$ is the Goldstino chiral superfield, 
which satisfies \cite{Rocek:1978nb,Lindstrom:1979kq,Casalbuoni:1988xh,Komargodski:2009rz} 
\be
\label{XNL1}
&&X_{NL}^2=0
\\
\label{XNL2}
&&\bar D^2 \bar X_{NL} - f + 2 \, {\cal C}\, X_{NL} =0 . 
\ee
These equations can be derived from the variation of 
\be
{\cal L} = \int d^4 \theta X \bar X + \lc \int d^2 \theta  \left( - f X + {\cal C} X^2 \right)+ c.c. \rc  
\ee
where $X$  is a chiral superfield and ${\cal C}$ is a chiral Lagrange multiplier superfield. 
In \cite{Farakos:2013zsa} it was shown that if $H$ satisfies these equations it also solves the
equation (\ref{eomH}). 
We therefore find that $H$ contains the Goldstino sector
that we found in (\ref{218}). 
Equations (\ref{XNL1}) and (\ref{XNL2}) are more restrictive than \eqref{chG1} and  \eqref{chG2}  
since they also lead to equations of motion for the Goldstino component. 
Indeed, equation (\ref{eomH}) can be solved only in terms of  superfields which satisfy appropriate equations of motion, 
since it is itself an equation of motion. 
For further discussion on the $X_{NL}$ Goldstino superfield 
and applications to particle physics see 
\cite{Antoniadis:2010hs,Antoniadis:2011xi,Dudas:2011kt,
Dudas:2012fa,Antoniadis:2012ck,Farakos:2013ih,Goodsell:2014dia,Dudas:2015vka}. 
Also, relations between different Goldstino realizations were given in \cite{Kuzenko:2011tj}.

Apart from the Goldstino sector, 
there is also the massive sector.   
For the chiral superfields $\Phi$ and $Y$ we find 
\be
\label{YP1}
\bar D^2 \bar Y &=& m \Phi 
\\
\label{YP2}
\bar D^2 \bar \Phi  &=& m Y 
\ee
where we have used $D^2 \bar \S = D^2 Y$ which follows from (\ref{sol1}). 
Equations (\ref{YP1}) and (\ref{YP2}) describe a 
pair of massive chiral multiplets, 
with Dirac masses for the femionic sector, 
exactly as we found from the component discussion in (\ref{218}).

Let us see what happens at low energy. 
The IR limit also implies the formal limit 
\be
m \rightarrow \infty 
\ee 
which leads to the decoupling of the massive modes, 
and we can set them to their vacuum values  
\be
Y=0  \ , \  \Phi=0 .
\ee 
This decoupling can be also seen from the component form (\ref{218}).  
For the $H$ superfield we have seen that in the supersymmetric vacuum it trivially vanishes.  
In the supersymmetry breaking vacuum the $H$ superfield stays massless and does not decouple in the IR, 
it describes the Goldstino sector. 
Indeed, if we call the Goldstino field $G_\alpha$ we have
\be
G_\alpha = D_\alpha X_{NL}| =  D_\alpha H | = D_\alpha ( \S - \bar Y) | =  D_\alpha \S | = \lambda_\alpha 
\ee
and from the component form (\ref{218}), we can see that the fermion $\lambda_\alpha$ is the only 
field that will appear in the IR. 
In the next section we will revisit the low energy behavior of the theory 
using supercurrent methods \cite{Komargodski:2009rz}.

\subsection{Complex linear multiplet, supersymmetry breaking  and  mediation}

For the massless complex linear multiplet we have 
\be
\bar D^2 \S =0 
\ee
with components defined as in (\ref{Scomponents}). 
The supersymmetry breaking mechanism we now describe was introduced in \cite{Farakos:2013zsa}. 
The Lagrangian used to achieve this is 
\be
\label{L66}
{\cal L} = -\int d^4 \theta \ \bar \Sigma \Sigma 
+ \frac{1}{8 f^2}  \int d^4 \theta   D^\alpha \Sigma D_\alpha \Sigma 
\bar D^{\dot{\beta}} \bar \Sigma  \bar D_{\dot{\beta}} \bar \Sigma   
\ee
with equations of motion
\be
  D_{\alpha} \left( \S +  \frac{1}{4 f^2} \bar D^{\dot\alpha}  \left( \bar D_{\dot\alpha}\bar \S 
 D^{\alpha} \S D_{\alpha} \S \right) \right)=0 
\ee
which integrates to
\be
\label{EOMS}
\S +  \frac{1}{4 f^2} \bar D^{\dot\alpha}  \left( \bar D_{\dot\alpha}\bar \S 
 D^{\alpha} \S D_{\alpha} \S \right) = \bar\Phi
\ee
where $\bar\Phi$ is an arbitrary chiral superfield zero mode of the $D_\alpha$ operator, 
and for consistency of (\ref{EOMS})  has to satisfy 
\be
\label{EQPHI}
\bar D^2 \bar \Phi = 0  . 
\ee 
Similarly to the previous model there is a supersymmetric vacuum in which
\be
\S = \bar\Phi
\ee
and the theory just reduces to a free chiral superfield. There is also a supersymmetry breaking vacuum solution
in which we solve the equation 
using the same reasoning as when solving equation (\ref{eomH}) leading to
\be
\label{SOL2}
\S = \bar \Phi  +  X_{NL}  
\ee 
with $X_{NL}$ satisfying (\ref{XNL1}) and (\ref{XNL2}) and $\Phi$ satisfying (\ref{EQPHI}). 
In the supersymmetric vacuum, the theory is described 
by a massless chiral superfield and in the SUSY breaking vacuum the theory contains a massless
chiral superfield and a massless Goldstino. All the excitations of the model stay massless and the only thing
that happens in the SUSY breaking vacuum is that there is a new propagating fermionic degree of freedom, the
Goldstino. 
For similar models with chiral superfields see for example 
\cite{Cecotti:1986jy,Buchbinder:1994iw,Buchbinder:1994xq,
Khoury:2010gb,Farakos:2012qu,Koehn:2012ar,Adam:2013awa,
Nitta:2014pwa,Nitta:2014fca,Nitta:2015uba,Ciupke:2015msa}. 
For supersymmetry breaking with a modified complex linear see \cite{Kuzenko:2011ti}.

A natural question to ask is how disentangled these degrees of freedom are. Since they all have the same mass they could
mix in some nontrivial way. To get a clearer picture of the independence of the degrees of freedom of the theory, we will now show how to mediate the supersymmetry breaking to the scalar sector.  
This can be achieved by modifying the higher derivative term
\be
\label{L6}
{\cal L} = -\int d^4 \theta \ \bar \Sigma \Sigma 
+ \frac{1}{8 f^2}  \int d^4 \theta \left( 1 - \frac{2 M^2}{ f^2} \S \bar \S \right)  D^\alpha \Sigma D_\alpha \Sigma 
\bar D^{\dot{\beta}} \bar \Sigma  \bar D_{\dot{\beta}} \bar \Sigma   
\ee
where the $M^2$ term is there to mediate the supersymmetry breaking to the scalar sector, 
by giving rise to masses.

To study the vacuum structure we write down the bosonic sector
\be
\nn
{\cal L}_\text{B} &=& \frac12 A \,  \p^{\alpha \dot \alpha} \p_{\alpha \dot \alpha} \bar A  - F \bar F 
+ P^{\alpha \dot \alpha}   \bar P_{\alpha \dot \alpha} 
\\
&& 
+  \frac{\left( 1 - \frac{2 M^2}{ f^2} A \bar A \right)}{2 f^2} \lc F^2 \bar F^2 
+  F \bar F P^{\alpha \dot \alpha}   \bar P_{\alpha \dot \alpha} 
+ \frac{1}{4} P^{\alpha \dot \alpha}    P_{\alpha \dot \alpha} \bar P^{\beta \dot \beta}   \bar P_{\beta \dot \beta} \rc. 
\ee
The equations for $P_{\alpha \dot \alpha}$ give $P_{\alpha \dot \alpha} = 0 $, 
and for the scalar $F$  we have two solutions 
\begin{enumerate}

\item The trivial vacuum with $\langle F \rangle=0$. 
Here supersymmety is not broken and $\langle V \rangle =0$ . 

\item The susy breaking vacuum with $\langle F \bar F \rangle= f^2$. 
Here supersymmetry is broken and $\langle V \rangle = \frac{f^2}{2}$ . 

\end{enumerate}

Let us study the supersymmetry breaking vacuum. 
If we expand the theory around that solution we see that the auxiliary fermion $\lambda_\alpha$  
now has a kinetic term
\be
- i \frac{\langle F \bar F \rangle}{f^2} \lambda_\beta \p^{\beta \dot \beta} \bar \lambda_{\dot \beta} = 
- i \lambda_\beta \p^{\beta \dot \beta} \bar \lambda_{\dot \beta} 
\ee
 and in fact is the Goldstone mode ($\delta \lambda_\alpha = f \,  \epsilon_\alpha + \cdots  $). 
The bosonic sector reads
\be
{\cal L}_\text{B} &=& \frac12 A \,  \p^{\alpha \dot \alpha} \p_{\alpha \dot \alpha} \bar A 
- \frac{f^2}{2} \frac{1}{1- \frac{2 M^2}{f^2} A \bar A}  
\ee
and for small field excitations the scalar potential becomes
\be
V = \frac{f^2}{2} \frac{1}{1- \frac{2 M^2}{f^2} A \bar A}  \simeq \frac{f^2}{2} + M^2 A \bar A 
\ee
therefore the scalar has become massive. 
One can check that the fermions $\psi_\alpha$ remain massless. 
Therefore, supersymmetry is broken and it is also mediated to the bosonic sector.

The superspace equations of motion which follow from the Lagrangian (\ref{L6}) are 
\be
\nn
\bar D^{\dot \gamma} \Big{\{ } \bar \S 
&+& \frac{1}{4 f^2}  D^\alpha \left( D_\alpha \S \bar D^{\dot \alpha} \bar \S \bar D_{\dot \alpha} \bar \S \right) 
\\
&+& \frac{M^2}{8 f^4} \bar \S D^\alpha \Sigma D_\alpha \Sigma 
\bar D^{\dot{\beta}} \bar \Sigma  \bar D_{\dot{\beta}} \bar \Sigma 
-  \frac{M^2}{4 f^4} D^\alpha \left( \S \bar \S D_\alpha \S \bar D^{\dot \alpha} \bar \S \bar D_{\dot \alpha} \bar \S \right)  
\Big{\} } = 0  . 
\ee
This equation can be rewritten as 
\be
\nn
\bar \S 
&+& \frac{1}{4 f^2}  D^\alpha \left( D_\alpha \S \bar D^{\dot \alpha} \bar \S \bar D_{\dot \alpha} \bar \S \right) 
\\
&+& \frac{M^2}{8 f^4} \bar \S D^\alpha \Sigma D_\alpha \Sigma 
\bar D^{\dot{\beta}} \bar \Sigma  \bar D_{\dot{\beta}} \bar \Sigma 
-  \frac{M^2}{4 f^4} D^\alpha \left( \S \bar \S D_\alpha \S \bar D^{\dot \alpha} \bar \S \bar D_{\dot \alpha} \bar \S \right) 
 =  \Phi  
\ee
where $\Phi$ is a chiral superfield arising as the zero mode of the $\bar D_{\dot\alpha}$ operator. 
If one is interested in the low energy behavior of the theory, 
the superspace equations have a simple solution as we will see.  
The low energy limit also implies the formal limit  
\be
M \rightarrow \infty . 
\ee
In this limit the equation breaks into two parts which decouple from each other.  
This happens because if we study the theory in the IR and $M \rightarrow \infty$, 
the  fluctuations of the fields are much smaller than $M$,  
therefore can not affect the $M$ dependent part; 
The two equations have to be solved independently.

For the part of the equations of motion  which does not contain $M$ we find  
\be
\bar \S 
+ \frac{1}{4 f^2}  D^\alpha \left( D_\alpha \S \bar D^{\dot \alpha} \bar \S \bar D_{\dot \alpha} \bar \S \right)  =  \Phi 
\ee
which is the same equation as in the model without mediation (\ref{EOMS}).
The solution is again the same, namely
\be
\label{248}
\bar \S = \bar X_{NL} + \Phi  
\ee
and
\be
\label{what}
D^2 \Phi = 0 
\ee
where the presence of $X_{NL}$ indicates that we are looking at the supersymmetry breaking solution. 
Again the Goldstino is the auxiliary field $\lambda_\alpha = D_\alpha \S | = D_\alpha X_{NL} | $, 
which becomes propagating when supersymmetry is broken.

The part proportional to  $M$ 
should rather be solved as a constraint  than as an equations of motion. 
Indeed we find that 
\be
\frac{M^2}{8 f^4} \bar \S D^\alpha \Sigma D_\alpha \Sigma 
\bar D^{\dot{\beta}} \bar \Sigma  \bar D_{\dot{\beta}} \bar \Sigma 
-  \frac{M^2}{4 f^4} D^\alpha \left( \S \bar \S D_\alpha \S \bar D^{\dot \alpha} \bar \S \bar D_{\dot \alpha} \bar \S \right) = 0
\ee
is always satisfied if 
we use (\ref{248}) and   constrain $\Phi$ to satisfy 
\be
\label{squarks}
 X_{NL} \, \Phi =0. 
\ee
It has been shown in \cite{Komargodski:2009rz}  that this particular constraint (\ref{squarks}) 
corresponds to the decoupling of the scalar lowest component of 
$\Phi$, namely $A$, 
which is replaced with Goldstino and $\psi$ fermions. 
Indeed, 
an inspection of the component form shows that in the limit $M \rightarrow \infty$, 
the scalar becomes very heavy and decouples from the IR physics. 
Now equation  (\ref{what}) together with (\ref{squarks})  makes perfect sense; 
it describes a massless fermion with no superpartner. 
This fermion is of course not the Goldstino since $\langle \bar D^2 \bar \Phi|\rangle =0 $.

Alternatively, one may mediate the supersymmetry breaking to the fermionic sector ($\psi_\alpha$) via the term
\be
\label{MPSI}
{\cal L}_{M_{\psi}} =
\frac{ M_{\psi} }{8 f^4}  \int d^4 \theta  
\left( \bar D^{\dot \gamma} \S \bar D_{\dot \gamma} \S + 
D^\gamma \bar \S D_\gamma \bar \S \right)
D^\alpha \Sigma D_\alpha \Sigma 
\bar D^{\dot{\beta}} \bar \Sigma  \bar D_{\dot{\beta}} \bar \Sigma   
\ee
which in the breaking vacuum generates masses 
\be
{\cal L}_{M_{\psi}}|_\text{broken vacuum} = \frac{M_{\psi}}{2} 
\left( \bar \psi^{\dot \gamma} \bar \psi_{\dot \gamma} + 
\psi^\gamma \psi_\gamma \right) + \cdots 
\ee
In the formal limit  
\be
M_{\psi} \rightarrow \infty 
\ee
which leads to the decoupling of the very massive fermion $\psi_\alpha$, 
the part of the equations of motion which is proportional to $M_{\psi}$ 
becomes a constraint and enforces the condition
\be
X_{NL} \bar D_{\dot \gamma} \bar \Phi  = 0 . 
\ee
This constraint has indeed been shown to correspond 
to the decoupling of the fermionic sector of matter superfields \cite{Komargodski:2009rz}. 
We therefore see that all the sectors except the Goldstino, 
can be consistently decoupled in the IR by introducing mass terms, 
leaving behind only the Goldstino superfield.

For the vacuum where supersymmetry is not broken we find 
\be
\S = \bar \Phi
\ee
with
\be
\bar D^2 \bar \Phi =0  
\ee
and no further constraints on $\Phi$.

\subsection{Comments on duality}

It is well known that
the  complex linear superfield can be dualized to a chiral superfield.
Similarly, the CNM multiplet is known to be dual to two massive chiral superfields.
The duality is not dependent on any special properties of the model such as the existence 
%of isometries
of the target space of a sigma model and therefore believed to be valid quite generally.
The procedure can be outlined as follows. Start with a theory defined by a Lagrangian depending on a complex
linear superfield and its derivatives
\be
%\int d^4x \; 
\int \! d^4\theta \, L(\S,\bar\S,D\bar\S,\bar D\S,\ldots).
\ee
We turn $\S$ into an unconstrained superfield by introducing a chiral field $\Phi$
\be\label{parent} 
%\int d^4x \; 
\int \! d^4\theta \left(L(\S,\bar\S,D\bar\S,\bar D\S,\ldots) + \Phi\S +\bar\Phi\bar\S\right).
\ee
Integrating out the chiral field $\Phi$ imposes the complex linearity constraint on $\S$ which gives back the original theory.
If we on the other hand integrate out $\S$ we get a complicated equation
\be\label{invert}
\Phi = -\frac{\partial L}{\partial \S} + \bar D\frac{\partial L}{\partial \bar D\S} + \ldots
\ee
which needs to be inverted as 
$\S = \S(\Phi,\bar\Phi,D\Phi,\bar D\bar\Phi,\ldots)$ and inserted back in (\ref{parent}) for us 
to be able to write the action of the dual 
theory depending on the chiral superfield $\Phi$. Thinking of the field $\S$ as a small fluctuation around a vacuum value
and organizing the right hand side of (\ref{invert}) in a series with smaller and smaller terms one may
invert the series term by term to find $\S$ as a function of $\Phi$. 
 
Although straightforward, this procedure becomes nontrivial when the theory has several possible vacua around
which we may invert the equations of motion (\ref{invert}). Choosing different vacua gives different dual theories
so the duality procedure, although valid, will capture only the physics of the particular vacua around which we choose
to invert. Also, if there are new propagating degrees of freedom in the vacuum at hand, the dual theory will not see
them since they belong to the background. One would have to insert them by hand after performing the duality.
We have already seen that by introducing superspace higher derivatives together with complex linear superfields or
CNM multiplets, we do get theories with several vacua. Let us look at how the duality works for several
interesting examples .

As an example we take the complex linear theory with the higher derivative term discussed in (\ref{L66}). 
The equation that needs to be inverted is then
\be\label{EOMagain}
\bar\Phi = \S +  \frac{1}{4 f^2} \bar D^{\dot\alpha}  \left( \bar D_{\dot\alpha}\bar \S 
 D^{\alpha} \S D_{\alpha} \S \right). 
\ee
Following the procedure outlined above we can now invert this relation around the two vacua of the theory
\be
\S &=& 0 + \dots \\
\S &=& X_{NL} + \ldots
\ee
To first order in $\bar\Phi$ we get
\be
\S &=& 0 + \bar\Phi +\dots \\
\S &=& X_{NL} + \bar\Phi +\ldots
\ee
When we insert this into (\ref{EOMagain}) to find the next order corrections we see that due to the particular structure
of the higher derivative term, we in fact have the full inverted solution in both cases. If we insert any of these solutions
into the original action we get a free chiral theory, the Goldstino of the supersymmetry breaking vacua needs to be inserted
by hand.

It is instructive to contrast this model with the very similar looking theory defined by the Lagrangian
\be
\label{6266}
{\cal L} =  - \int d^4 \theta \Sigma \bar \Sigma 
+ \frac{1}{8 f^2}  \int d^4 \theta  D^\alpha \bar \Sigma D_\alpha \bar \Sigma 
\bar D^{\dot{\beta}} \Sigma  \bar D_{\dot{\beta}} \Sigma  . 
\ee
Here the superspace equations of motion are 
\be
\label{d3}
\bar \Phi = \S+ \frac{1}{4 f^2} D^{\alpha} \left( D_{\alpha} \bar\S 
\bar D^{\dot\alpha} \S  \bar D_{\dot\alpha} \S \right) . 
\ee
In this case there is no supersymmetry breaking vacuum and the only possible solution is to invert around
the trivial vacuum $\S = 0$
\be
\S = 0 + \bar\Phi +\ldots
\ee
If we insert this into (\ref{d3}) we are left with term of third order in $\Phi$ so the procedure has to continue.
Following this program to the end one can shown that 
the equation (\ref{d3}) can be inverted as 
\be
\S = f(\Phi, D_\alpha \Phi, \bar \Phi, \cdots )
\ee 
and after plugging back into (\ref{6266}) one ends up with a higher derivative theory  
for the chiral superfield $\Phi$ \cite{Rocek:1997hi,GonzalezRey:1998kh}.

As we have seen, to perform the inversion procedure, 
we had to treat the new degrees of freedom as a background, 
therefore a Lagrangian description of the new degrees of freedom was not possible. 
Now we would like to present a complementary approach to the previous discussion, 
which will allow us to find  a Lagrangian which will also include the new superfields. 

Let us  remind the reader the duality for the free complex linear multiplet.  
We have  
\be
{\cal L} &=& -\int d^4 \theta \ \bar \Sigma \Sigma 
 + \int d^4 \theta  \ \Phi \S  + \int d^4 \theta \ \bar \Phi \bar \S  
\ee
where $\Phi$ is a  chiral superfield but $\S$ is unconstrained.  
By integrating out $\Phi$ we get that $\bar D^2 \S = 0$, 
therefore we have a complex linear multiplet. 
If we now define the unconstrained superfields $\Xi$ as 
\be
\Xi = \S - \bar \Phi 
\ee 
the theory becomes
\be
\label{GPhi}
{\cal L} &=&  \int d^4 \theta \ \bar \Phi \Phi 
 -\int d^4 \theta \ \bar \Xi \Xi . 
\ee
We may trivially integrate out $\Xi$. 
We see that the theory is dual to a free massless chiral superfield. 
The last step completes the duality and is important  for  our discussions. 
We note that if one turns to the component form of (\ref{GPhi}), 
there will not be any kinetic terms for the component fields of $\Xi$, 
therefore here it is indeed non-dynamical.

The general complex linear model can be written as 
\be
{\cal L} = - \int d^4 \theta \, \S \bar \S 
+ \int d^4 \theta \, \Omega (\S \, , \bar \S)  
+ \int d^4 \theta  \ \Phi \S  + \int d^4 \theta \ \bar \Phi \bar \S  
\ee 
where $\Omega (\S \, , \bar \S)$ may contain also 
superspace higher derivative terms ($D^\alpha \S \, , D^2 \S \cdots$) as we said earlier. 
Here $\S$ is unconstrained but becomes a complex linear when we integrate out the chiral superfield $\Phi$. 
Again we define 
\be
\Xi = \S - \bar \Phi 
\ee
and the theory becomes
\be
\label{full-1}
{\cal L} =  \int d^4 \theta \ \bar \Phi \Phi    -\int d^4 \theta \ \bar \Xi \Xi 
+ \int d^4 \theta \, \Omega (\Xi + \bar \Phi \, , \bar \Xi + \Phi) . 
\ee 
Now we have to complete the duality by integrating out $\Xi$ from (\ref{full-1}).  
Two things may happen here. 
\begin{enumerate}

\item The variation with respect to $\Xi$ yields algebraic equations and $\Xi$ can be integrated out. 
In case the equations are complicated but solvable, 
a   solution can still  be found around $\Xi = 0$, 
up to the desired order, by inverting them as  described earlier. 

\item The variation with respect to $\Xi$ yields  equations of motion and $\Xi$ can not be integrated out; 
$\Xi$ is dynamical and it is related to dynamical auxiliary fields. 
The Lagrangian \eqref{full-1} makes the new degrees of freedom manifest, 
and it provides the Lagrangian description for the theory.

\end{enumerate}
If a theory is described by the first or the second case depends on the particular  form of the superspace function  
$\Omega (\S \, , \bar \S \, , D^\alpha \S \, , D^2 \S \cdots) $. 
To clarify our discussion we will study two explicit cases where $\Omega$ does contain superspace higher derivatives. 
The superspace higher derivatives we introduce here, 
have the property to give rise to kinetic terms for the auxiliary fields in the component form.   
We will see that this is related to $\Xi$ being dynamical.

First we introduce the model which we saw that gives rise to the kinetic terms for the 
auxiliary fermion in the broken vacuum. 
We have
\be
{\cal L} = - \int d^4 \theta \S \bar \S 
+ \frac{1}{8 f^2}  \int d^4 \theta  D^\alpha \Sigma D_\alpha \Sigma 
\bar D^{\dot{\beta}} \bar \Sigma  \bar D_{\dot{\beta}} \bar \Sigma 
 + \int d^4 \theta \,  \S  \Phi  + \int d^4 \bar \theta \, \bar \S  \bar \Phi
\ee 
which with the definition
\be
\Xi = \S - \bar \Phi 
\ee
becomes  
\be
\label{full}
{\cal L} =  \int d^4 \theta \ \bar \Phi \Phi 
  -\int d^4 \theta \ \bar \Xi \Xi 
+ \frac{1}{8 f^2}  \int d^4 \theta  D^\alpha \Xi D_\alpha \Xi 
\bar D^{\dot{\beta}} \bar \Xi  \bar D_{\dot{\beta}} \bar \Xi 
\ee 
with $\Xi$ unconstrained. 
Notice that the chiral sector and the $\Xi$ sector have completely decoupled. 
To complete the duality procedure one integrates out $\Xi$. 
The variation with respect to $\Xi$ yields
\be
\label{Geq1}
\bar \Xi + \frac{1}{4 f^2} D^\alpha \left( D_\alpha \Xi \bar D^{\dot \alpha} \bar \Xi \bar D_{\dot \alpha}  \bar \Xi \right) =0 . 
\ee 
Equation (\ref{Geq1}) has two solutions. 
The first solution 
\be
\Xi=0 
\ee 
represents the theory around the supersymmetry preserving vacuum. 
Notice that in this vacuum the free  theory remains intact, 
exactly as we found for the component sector (\ref{intact}), 
and in particular it reads 
\be
\label{full777}
{\cal L} =  \int d^4 \theta \ \bar \Phi \Phi . 
\ee 
The second solution is  
\be
\Xi = X_{NL} . 
\ee 
This solution requires new degrees of freedom (a Goldstino in particular), 
and would not be captured by expanding around $\Xi=0$. 
This is related to the fact that equation (\ref{Geq1}) contains dynamics in the broken vacuum. 
This can be also seen in the component level, 
where expanding around $\langle D^2 \Xi | \rangle \ne 0 $  (the supersymmetry breaking vacuum) 
one finds there is a propagating Goldstino mode, which is a component field of $\Xi$. 
The Lagrangian for this vacuum is therefore  \eqref{full}.

We can also look at  an example with a complex linear superfield but with no supersymmetry breaking 
\be
{\cal L} =  - \int d^4 \theta \Sigma \bar \Sigma +  \alpha  \int d^4 \theta  D^2 \Sigma \bar D^2 \bar \Sigma . 
\ee 
The bosonic sector of this theory contains 4 real additional propagating 
bosonic modes ($F$ and $\p^{\alpha \dot \alpha} P_{\alpha \dot \alpha}$) 
which form an on-shell supermultiplet with the auxiliary spinors $\chi_\alpha$ and $\lambda_\alpha$, 
which also become propagating and together form a massive Dirac spinor. 
There are no ghosts for $\alpha > 0$ which we will assume henceforth. 
If we start the duality procedure we have
\be
{\cal L} =  - \int d^4 \theta \Sigma \bar \Sigma 
+  \alpha  \int d^4 \theta  D^2 \Sigma \bar D^2 \bar \Sigma  
+ \int d^4 \theta \ls \Phi \S + \bar \Phi \bar \S \rs 
\ee
for $\Phi$ chiral and $\S$ unconstrained. 
Now we define 
\be
\Xi = \S - \bar \Phi 
\ee
and we have 
\be
\label{secondex}
{\cal L} =  \int d^4 \theta   \Phi \bar \Phi  - \int d^4 \theta \Xi \bar \Xi 
+ \alpha \int d^4 \theta  D^2 \Xi \bar D^2 \bar \Xi . 
\ee 
The equations of motion for $\Xi$ are  
\be
\label{G4D}
\alpha D^2 \bar D^2 \bar \Xi = \bar \Xi . 
\ee 
Taking a small $\alpha$, and inverting around $\Xi = 0$ one would find that $\Xi$ should vanish, 
which clearly does not represent all the propagating degrees of freedom. 
In fact (\ref{G4D}) is a dynamical equation which 
describes two massive chiral superfields and therefore $\Xi$ can not be integrated out, 
and the Lagrangian description of the theory is precisely \eqref{secondex}. 
To explain the origin of the propagating chiral superfields, 
we can rewrite the model as
\be
\label{888}
\begin{split}
{\cal L} =&  \int d^4 \theta   \Phi \bar \Phi  - \int d^4 \theta \Xi \bar \Xi 
+ \alpha \int d^4 \theta  \bar S S  
\\
& + \int d^2 \theta \, T \left( S - \bar D^2 \bar \Xi \right) 
+ \int d^2 \theta \, \bar T \left( \bar S - D^2 \Xi \right) 
\end{split}
\ee
where $T$ is a chiral Lagrange multiplier. 
After we integrate out $\Xi$
(which now has equations $\Xi = - T$) and rescale $S$ 
with $\sqrt \alpha$, the Lagrangian \eqref{888} becomes 
\be
{\cal L} =  \int d^4 \theta   \Phi \bar \Phi  + \int d^4 \theta T \bar T 
+ \int d^4 \theta  \bar S S  
+ \frac{1}{\sqrt{\alpha}} \int d^2 \theta \, T  S  
+ \frac{1}{\sqrt{\alpha}} \int d^2 \theta \, \bar T  \bar S . 
\ee 
We see that the new modes have mass $1/ \sqrt{\alpha}$, 
therefore if we had performed the inversion for small $\alpha$ 
we would be effectively decoupling them.

A similar situation appears  in the supergravity theory. 
The duality between the new-minimal and the old-minimal formulations can be understood as a duality \cite{Lindstrom:1983rt} 
between the chiral and the real linear   compensator 
which when gauge fixed, 
break the superconformal theory to superPoincare \cite{Ferrara:1983dh}. 
The supergravity-matter theories with no curvature higher derivatives can be in principle dualized to each other, 
but the chiral-linear duality does not offer a complete 
description when higher curvature terms are present \cite{Cecotti:1987qe}. 
In that case the compensator equations seize to be algebraic 
and it may not be integrated out in order to lead the  dual theory.  
In these cases the equivalent theory contains the gravitational sector but also 
additional propagating sectors appear. 
In the component form, one can see this by the fact that in higher curvature supergravity 
some of the auxiliary fields become propagating \cite{Cecotti:1987qe,Cecotti:1987sa,Farakos:2015hfa}.

\section{Supercurrents  and low energy limits}

In this section we study  models of non-minimal superfields, 
calculate their supercurrent when they include superspace higher derivatives, 
and study the IR limits. 
For the supersymmetry breaking vacua we give low energy descriptions.

The supercurrent conservation equations (which hold only when one uses the equations of motion) have the generic form  \cite{Ferrara:1974pz,Magro:2001aj,Shizuya:1986xt,Osborn:1998qu,Clark:1988es,
Komargodski:2010rb,Kuzenko:2010am,Kuzenko:2010ni,Arnold:2012yi,Clark:1995bg} 
\be
\label{se}
\bar D^{\dot \alpha} {\cal J}_{\alpha \dot \alpha} = {\cal Y}_{\alpha} +  {\cal X}_{\alpha} 
\ee
where the supercurrent ${\cal J}_{\alpha \dot \alpha}$ is a real superfield, 
and the superfields ${\cal Y}_{\alpha}$ and ${\cal X}_{\alpha}$ satisfy 
\be
\label{chi1}
\bar D_{\dot \alpha} {\cal X}_{\alpha}  &=& 0 
\\
\label{chi2}
D^{\alpha} {\cal X}_{\alpha} +  \bar D^{\dot \alpha} \bar {\cal X}_{\dot \alpha}  &=& 0 
\ee
and 
\be
\label{y1}
\bar D^2 {\cal Y}_{\alpha}  &=& 0 
\\
\label{y2}
D_{\alpha} {\cal Y}_{\beta} + D_{\beta} {\cal Y}_{\alpha}   &=& 0 . 
\ee 
The superfields which enter the right hand side of the supercurrent equation (\ref{se}), 
have IR properties related to supersymmetry breaking. 
{}From the identities for ${\cal Y}_\alpha$ we see that locally it can always be written as
\be
{\cal Y}_\alpha = D_\alpha X 
\ee
where
\be
\bar D_{\dot \alpha} X = 0 . 
\ee 
It was pointed out in \cite{Komargodski:2009rz} that  when supersymmetry is broken, 
one will find (on-shell) for the low energy 
\be
X \rightarrow X_{IR} =   X_{NL} . 
\ee 
We will show now that this property holds also for the models 
of complex linear and CNM multiplets which break supersymmetry 
with superspace higher derivatives.

\subsection{Supercurrents from the Noether procedure}

First we have to identify the supercurrents, 
which is done by turning to the Noether procedure \cite{Magro:2001aj,Shizuya:1986xt,Osborn:1998qu}. 
A superdiffeomorphism of a superfield $S$ is given by the transformation 
\be
S \rightarrow e^{i \Delta} S e^{-i \Delta}
\ee 
where 
\be
\Delta = \Delta^\alpha D_\alpha +  \Delta^{\dot \alpha} \bar D_{\dot \alpha} 
+ \Delta^{\alpha \dot \alpha} \p_{\alpha \dot \alpha} .   
\ee
If the superfield $S$ is  
a complex linear ($\bar D^2 S =0$), 
this property has to be preserved by the superdiffeomophism
which therefore leads to the restrictions 
\be
\nn
\bar D_{\dot \alpha} \Delta^\alpha  &=& 0 
\\
\nn
i \delta^{\dot \alpha}_{\dot \beta} \Delta^\alpha  &=& \bar D_{\dot \beta}\Delta^{\alpha \dot \alpha}
\\
\bar D^2 \Delta^{\alpha \dot \alpha} &=& 0 
\\
\nn
\bar D^2 \Delta^{\dot \alpha} &=& 0 
\ee
which are solved by
\be
\nn
\Delta^{\alpha \dot \alpha} &=& \bar D^{\dot \alpha} L^\alpha 
\\
\Delta^\alpha &=& i \bar D^2 L^\alpha 
\\
\nn
\Delta^{\dot \alpha} &=& \bar D_{\dot \beta} L^{\dot \beta \dot \alpha}  
\ee 
with $L^\alpha$ and $L^{\dot \beta \dot \alpha}$ both complex and unconstrained. 
%Indeed, in this case we 
%have 
%\be
%\begin{split}
%[ \bar D^{\dot \rho} , \Delta ] \, {\cal O} =& 
%\frac12 \left( \bar D \bar D L^{\dot  \rho \dot \beta} \right) \bar D_{\dot \beta} {\cal O}
%\\
%[ \bar D \bar D, \Delta ] \, {\cal O} =& 
%- \left( \Delta + \frac12 \bar D \bar D L^{\dot \rho \dot \beta} C_{\dot \beta \dot \rho} \right) \left( \bar D \bar D {\cal O} \right)   
%\end{split}
%\ee
It is straightforward to check that this choice of parameters is also compatible with $S$ being chiral 
($\bar D_{\dot \alpha} S = 0$) and also with the CNM multiplet.

An infinitesimal transformation for the complex linear is
\be
\delta_\text{superdiff} \S = [i \Delta , \S] = - \bar D^2 L^\alpha D_\alpha \S 
+ i \bar D^{\dot \alpha} L^{\alpha} \, \p_{\alpha \dot \alpha} \S 
+ i  \bar D_{\dot \beta} L^{\dot \beta \dot \alpha} \, \bar D_{\dot \alpha} \S . 
\ee
The superspace Noether procedure \cite{Magro:2001aj,Shizuya:1986xt,Osborn:1998qu} 
then directly gives conserved complex currents
\be
 \delta {\cal L} 
= - \int d^4 \theta \left( \bar D^{\dot \alpha} L^\alpha  {\cal J}_{\alpha \dot \alpha}
+ \bar D_{\dot \beta} L^{\dot \beta \dot \alpha}  {\cal J}_{\dot \alpha} \right) + c.c. 
\ee
with conservation equations 
\be
\bar D^{\dot \alpha} {\cal J}_{\alpha \dot \alpha} = 0
\ee 
and 
\be
\bar D_{\dot \beta} {\cal J}_{\dot \alpha} = 0
\ee 
but generically 
\be
D^{\alpha} {\cal J}_{\alpha \dot \alpha} \ne 0
\ee 
since ${\cal J}_{\alpha \dot \alpha}$ is not necessarily real.

We may now use improvement terms to bring ${\cal J}_{\alpha\dot\alpha}$, ${\cal X}_\alpha$ and ${\cal Y}_{\alpha}$ 
to the desired form (\ref{se}). 
This can be  done by using shifts which change the form of the supercurrent ${\cal J}_{\alpha \dot \alpha}$, 
at the same time as they also change  ${\cal X}_{\alpha}$ and ${\cal Y}_{\alpha} $. 
These shifts can be found in Table 1.
Furthermore, since in the variation of the action, ${\cal J}_{\alpha\dot\alpha}$ is 
multiplied with $\bar{D}^{\dot\alpha} L^\alpha$, we can interchange a term in ${\cal J}_{\alpha\dot\alpha}$ of 
the form   $\bar D_{\dot \alpha} X_{\ \alpha \dot \beta }^{\dot \beta}$   
with  $-2 \bar D^{\dot \beta} X_{\dot \beta \alpha \dot \alpha}$.

\begin{table}[htb] \label{tabb2} 
%{\small
\begin{center}
$$
\begin{array}{|r|r|r|r|r|}\hline 
\text{\bf Shifts } 
&  
\text{\bf Type A } 
 \quad 
 \quad  
 & 
\text{\bf Type B } 
 \quad \quad 
 & 
\text{\bf Type C } 
 \quad \quad 
 & 
\text{\bf Type D } 
 \quad \quad 
 \\
\hline 
\hline
\quad
{\cal J}_{\alpha\dot\alpha}  
\rightarrow 
%\quad 
\quad
&  
\ 
{\cal J}_{\alpha\dot\alpha} 
+ [D_{\alpha},\bar{D}_{\dot\alpha}] U \ 
&
\ 
{\cal J}_{\alpha\dot\alpha} 
+ i\partial_{\alpha\dot\alpha} U \ \ \ 
& 
\ 
{\cal J}_{\alpha\dot\alpha} 
+ \bar{D}_{\dot\alpha}D_{\alpha} U \ \ 
& 
\ 
{\cal J}_{\alpha\dot\alpha} 
+D_{\alpha}\bar{D}_{\dot\alpha}U \ \ 
\\
\hline
{\cal X}_{\alpha} 
\rightarrow 
%\quad  
\quad 
&  
\ 
{\cal X}_{\alpha} 
- 3 \bar{D}^2D_\alpha U \ \ 
&  
\ 
{\cal X}_{\alpha} 
+ \bar{D}^2D_\alpha U \ \ 
&  
\  
{\cal X}_{\alpha} 
+ 2\bar{D}^2D_\alpha U \ \ 
&  
\  
{\cal X}_{\alpha} 
- \bar{D}^2D_\alpha U \  \ \ 
\\
\hline
{\cal Y}_{\alpha} 
\rightarrow 
%\quad  
\quad 
&    
\ 
{\cal Y}_\alpha 
- D_\alpha \bar{D}^2 U \ \ \ 
&  
\ 
{\cal Y}_\alpha 
- D_\alpha \bar{D}^2 U \ \ 
&    
\ 
 {\cal Y}_\alpha 
 \quad \quad \quad
&   
\ 
{\cal Y}_\alpha 
- D_\alpha \bar{D}^2 U \ \ \  
\\
\hline
\end{array}
$$
\end{center}
\caption{\small The table presents the various shifts which can be used to bring the current to the desired form.} 
%}
\end{table}

We now proceed as follows. We start with a complex current ${\cal J}_{\alpha\dot\alpha}$ and
${\cal X}_\alpha = {\cal Y}_\alpha = 0$. We use all possible shifts and rewritings to make the current real. This
produces nonzero ${\cal X}_\alpha$ and ${\cal Y}_\alpha$, however, ${\cal X}_\alpha$ might not fulfil (\ref{chi2}).
To try to improve this, we may only perform shifts that respect the reality of ${\cal J}_{\alpha\dot\alpha}$. Those are given by
type A shifts with a real $U$ and type B shifts with an imaginary $U$. Finally we are left with a system
\be
\bar D^{\dot\alpha}{\cal J}_{\alpha\dot\alpha}={\cal X}_{\alpha}+{\cal Y}_\alpha
\ee
satisfying all the requirements. We may still perform shifts of type A with a real $U$ to change the system into the 
FZ-multiplet (${\cal X}_\alpha = 0$) or to the ${\cal R}$-multiplet (${\cal Y}_\alpha = 0$). 

%Let us  clarify the properties of the various shifts. 
%Firstly, the shifts always keep the properties of the  
%equations (\ref{chi1}), (\ref{y1}) and (\ref{y2}). 
%Secondly, 
%not all of these  transformations will preserve the reality properties of  ${\cal J}_{\alpha \dot \alpha}$ 
%or the property (\ref{chi2}) of ${\cal X}_\alpha$ for a general $U$. 
%Thirdly, only  the type A transformation for real $U$ will keep 
%the structure of the supercurrent equations  (\ref{se}) and 
%(\ref{chi2})  unchanged. 
%Therefore, one may use all  shifts  of Table 1 to make ${\cal J}_{\alpha \dot \alpha}$  real, 
%with $U$ not necessary real. 
%After this, 
%equation (\ref{chi2}) will in general not be satisfied, 
%but by using type B shift with pure imaginary $U$ 
%one can make ${X}_\alpha$ satisfy (\ref{chi2}). 
%In the end of this procedure, 
%the supercurrent (${\cal J}_{\alpha \dot \alpha}$) 
%will be real and equations (\ref{chi1}) to (\ref{y2}) will be satisfied. 
%Now one may use type A with real $U$ to bring the supercurrent 
%to the form of the FZ-multiplet or the ${\cal R}$-multiplet (if the theory is R-symmetric). 

In the next part of this section we will use the above methods to find the 
appropriate form of the supercurrents for the various cases, identify $X$, 
and study its IR flow.

%
%
%First, using a superfield $U$, 
%we can redefine the fields as 
%\be
%\label{UU}
%{\cal J}_{\alpha\dot\alpha} &\rightarrow & {\cal J}_{\alpha\dot\alpha} + [D_{\alpha},\bar{D}_{\dot\alpha}] U
%\nn
%\\
%{\cal X}_{\alpha} &\rightarrow& {\cal X}_{\alpha} - 3 \bar{D}^2D_\alpha U
%\\
%\nn
%{\cal Y}_\alpha &\rightarrow& {\cal Y}_\alpha - D_\alpha \bar{D}^2 U . 
%\ee
%Notice that this redefinition keeps for a real $U$
%
%
%or
%\be
%\nn
%{\cal J}_{\alpha\dot\alpha} &\rightarrow & {\cal J}_{\alpha\dot\alpha} + i\partial_{\alpha\dot\alpha} U
%\\
%{\cal X}_{\alpha} &\rightarrow& {\cal X}_{\alpha} + \bar{D}^2D_\alpha U
%\\
%\nn
%{\cal Y}_\alpha &\rightarrow& {\cal Y}_\alpha - D_\alpha \bar{D}^2 U
%\ee
%or
%
%
%
%
%\be
%\nn
%{\cal J}_{\alpha\dot\alpha} &\rightarrow & {\cal J}_{\alpha\dot\alpha} + \bar{D}_{\dot\alpha}D_{\alpha} U
%\\
%{\cal X}_{\alpha} &\rightarrow& {\cal X}_{\alpha} + 2\bar{D}^2D_\alpha U
%\\
%\nn
%{\cal Y}_\alpha &\rightarrow& {\cal Y}_\alpha
%\ee
%or
%
%
%
%\be
%\nn
%{\cal J}_{\alpha\dot\alpha} &\rightarrow & {\cal J}_{\alpha\dot\alpha} +D_{\alpha}\bar{D}_{\dot\alpha}U
%\\
%{\cal X}_{\alpha} &\rightarrow& {\cal X}_{\alpha} - \bar{D}^2D_\alpha U
%\\
%\nn
%{\cal Y}_\alpha &\rightarrow& {\cal Y}_\alpha - D_\alpha \bar{D}^2 U . 
%\ee
%

\subsection{IR limits of supersymmetry breaking vacua }

For the model of the  complex linear of \cite{Farakos:2013zsa} 
\be
{\cal L} = - \int d^4 \theta \S \bar \S 
+ \frac{1}{8 f^2}  \int d^4 \theta  D^\alpha \Sigma D_\alpha \Sigma 
\bar D^{\dot{\beta}} \bar \Sigma  \bar D_{\dot{\beta}} \bar \Sigma 
\ee
(with $\bar D^2 \S=0$) the Noether procedure gives
\be
\label{complexJ}
{\cal J}_{\alpha \dot \alpha} &=& -\frac12 \bar D_{\dot \alpha} \left( D_\alpha \S \, \bar Z \right) 
+ i \p_{\alpha \dot \alpha} \S \, \bar Z 
\\
{\cal J}_{\dot \alpha} &=& i \bar D_{\dot \alpha} \S \, \bar Z 
\ee
where
\be
\label{masslessZ}
Z = \S + \frac{1}{4 f^2} \bar D^{\dot \alpha} \left( \bar D_{\dot \alpha} \bar \S  
D^\alpha \S D_\alpha \S  \right) 
\ee
and the equations of motion are
\be
\label{Zcond1}
D_\alpha Z =0 . 
\ee
Note that $Z$ also satisfies
\be
\label{Zcond2}
\bar D^2 Z = 0 . 
\ee
It is easy to check that on-shell $\bar D^{\dot \alpha} {\cal J}_{\alpha \dot \alpha}=0 $  and also see that the ${\cal J}_{\alpha \dot \alpha}$ 
current is not real. 
To make the current real we use a combination of shift from table 1 to shift ${\cal J}_{\alpha\dot\alpha}$ with
\be
\label{325shift}
\frac12 \bar{D}_{\dot\alpha}D_{\alpha}(-Z\bar\Sigma+\Sigma\bar\Sigma - 2 T) 
+ \frac i2 \p_{\alpha \dot \alpha}(-\Sigma\bar\Sigma + Z \bar \S - \bar Z \S + T) 
\ee
where
\be
T &=& \frac{1}{2 f^2} (D\Sigma)^2(\bar{D}\bar\Sigma)^2 . 
\ee
If we also define
\be
\label{YL2}
\begin{split}
T_\beta &= \frac{1}{2 f^2} D_\beta \Sigma (\bar{D}\bar\Sigma)^2
\\
\bar{T}_{\dot\beta} &= \frac{1}{2 f^2} \bar{D}_{\dot\beta}\bar\Sigma (D\Sigma)^2  
\end{split}
\ee
we can write the resulting system as
\be
{\cal J}_{\alpha\dot\alpha} &=& - \frac 12 i \Sigma \partial_{\alpha\dot\alpha} \bar Z 
+ \frac12  i \bar\Sigma \partial_{\alpha\dot\alpha} Z 
+ \frac12 D^\beta(i\partial_{\alpha\dot\alpha}\Sigma \, T_\beta)
-\frac12 \bar{D}^{\dot\beta}(i\partial_{\alpha\dot\alpha}\bar\Sigma \, \bar{T}_{\dot\beta})
\\
\label{XXXX}
{\cal X}_\alpha &=& \frac12 \bar{D}^2D_{\alpha}(\Sigma\bar\Sigma - 3 T - Z \bar \S - \bar Z \S) 
\\
\label{YYYY}
{\cal Y}_{\alpha} &=& \frac12 D_\alpha\bar{D}^2(\Sigma\bar\Sigma - T - Z \bar \S - \bar Z \S) . 
\ee
Now ${\cal J}_{\alpha \dot \alpha}$ has become real 
and  
${\cal X}_\alpha$  satisfies \eqref{chi2}. 
Notice that in (\ref{YYYY}) after the shift (\ref{325shift}) the last term  
will appear like $+ \bar Z \S$, 
but by using the equations of motion this term will vanish 
due to the fact that $\bar D^2$ acts on it, 
therefore one may flip the sign to bring it in the form of (\ref{YYYY}) as we have done here, 
such that everything inside (\ref{YYYY}) is real.

We are still allowed to add improvement terms 
to bring the supercurrent equation to the desired form either of the 
FZ-multiplet or the ${\cal R}$-multiplet. 
But to keep the reality properties of ${\cal J}_{\alpha\dot\alpha}$ 
and the properties (\ref{chi2}) of ${\cal X}_\alpha$, we can 
only use the type A shift with a real $U$. 
To find the ${\cal R}$-multiplet we perform a type A shift with 
\be
U = \frac12 \left(  \Sigma\bar\Sigma - T - Z \bar \S - \bar Z \S  \right)
\ee
which gives a real ${\cal J}_{\alpha \dot \alpha}$  and 
\be
\begin{split}
{\cal X}_\alpha &= \bar D^2 D_\alpha \ls  - \S \bar \S + Z \bar \S + \bar Z \S  \rs 
\\
X &= 0 
\end{split}
\ee 
which together with the new supercurrent satisfy  
\be
\bar D^{\dot \alpha} {\cal J}_{\alpha \dot \alpha} = {\cal X}_\alpha . 
\ee
The fact that we can bring the supercurrent conservation equation in this form 
shows that this model can be coupled to the new-minimal supergravity consistently.

Now we turn to the FZ-multiplet. 
By performing a type A shift with 
\be
U= \frac16 (\Sigma\bar\Sigma - 3 T - Z \bar \S - \bar Z \S)
\ee 
we get a system with a real ${\cal J}_{\alpha \dot \alpha}$ and 
\be
\begin{split}
{\cal X}_\alpha &= 0 
\\
\label{Yold}
{\cal Y}_\alpha &= \frac23 D_\alpha \bar D^2 T   . 
\end{split}
\ee
{}From (\ref{Yold}) we find 
\be
\label{XXX}
X = \frac{2}{3} \bar D^2 T . 
\ee 
The new supercurrent and $X$ satisfy  
\be
\bar D^{\dot \alpha} {\cal J}_{\alpha \dot \alpha} = D_\alpha X  
\ee
which shows that this model can be also coupled to the old-minimal supergravity.

Now we want to study the IR limit of $X$ for the supersymmetry breaking vacuum. 
We have found that $\S = X_{NL} + \bar \Phi$, 
therefore we insert this in the expression for $X$ (\ref{XXX}) to find 
\be
X =   \frac{1}{3 f^2} \bar D^2 \ls (D X_{NL})^2(\bar{D}\bar X_{NL})^2 \rs 
\ee
which gives
\be
X = \frac{1}{3} f  X_{NL} . 
\ee
We see that $X$ for the supersymmetry breaking vacuum is proportional to $X_{NL}$, 
and this will also hold in the IR. 
Therefore we confirm that $X$ flow to $X_{NL}$ in the IR as was advocated in \cite{Komargodski:2009rz}.

For the CNM Lagrangian (\ref{L1}) we find from the Noether procedure 
\be
{\cal J}_{\alpha \dot \alpha} =  
 -\frac12 \bar D_{\dot \alpha} \left( D_\alpha \S \, \bar Z \right) 
+ i \p_{\alpha \dot \alpha} \S \, \bar Z 
+ \frac{1}{2} \bar D_{\dot \alpha} (\bar \Phi D_\alpha \Phi ) 
- i \bar \Phi \p_{\alpha \dot \alpha} \Phi 
\ee
with
\be
Z = \S + \frac{1}{4 f^2} \bar D^{\dot \alpha} \left( \bar D_{\dot \alpha} \bar \S  
D^\alpha \S D_\alpha \S  \right)   
\ee 
and the equations of motion are
\be
\bar Z = Y \ , \ \bar D^2 \bar Y = m \Phi \ , \ \bar D^2 \bar \Phi = m Y .  
\ee
The only difference from the massless complex linear model is the presence of the chiral superfield $\Phi$ 
inside the supercurrent. 
To bring ${\cal J}_{\alpha \dot \alpha}$ to the desired form we shift with 
\be
\frac12 \bar{D}_{\dot\alpha}D_{\alpha}(-Z\bar\Sigma+\Sigma\bar\Sigma - 2 T) 
+ \frac i2 \p_{\alpha \dot \alpha}(-\Sigma\bar\Sigma + Z \bar \S - \bar Z \S + T) 
+ \frac{i}{4} \p_{\alpha \dot \alpha}(\Phi \bar \Phi)
\ee
which is a combination of the various types of shifts shown in table 1. 
After this we find 
\be 
{\cal J}_{\alpha\dot\alpha} &=& - \frac 12 i \Sigma \partial_{\alpha\dot\alpha} \bar Z 
+ \frac12  i \bar\Sigma \partial_{\alpha\dot\alpha} Z 
+ \frac12 D^\beta(i\partial_{\alpha\dot\alpha}\Sigma \, T_\beta)
-\frac12 \bar{D}^{\dot\beta}(i\partial_{\alpha\dot\alpha}\bar\Sigma \, \bar{T}_{\dot\beta}) 
\\
\nn
&& + \frac{1}{2} \bar D_{\dot \alpha} \bar \Phi  D_\alpha \Phi  
- \frac{i}{4} \bar \Phi \p_{\alpha \dot \alpha} \Phi 
+ \frac{i}{4} \Phi \p_{\alpha \dot \alpha} \bar \Phi 
\\
\label{XXXXxx}
{\cal X}_\alpha &=& \frac12 \bar{D}^2D_{\alpha} \ls \Sigma\bar\Sigma - 3 T - Z \bar \S - \bar Z \S   + \frac12 \Phi \bar \Phi \rs  
\\
\label{YYYYyy}
{\cal Y}_{\alpha} &=& \frac12 D_\alpha\bar{D}^2 \ls  \Sigma\bar\Sigma - T - Z \bar \S + \bar Z \S  - \frac12 \Phi \bar \Phi  \rs
\ee
where $T$, $T_\alpha$ and $\bar T_{\dot \alpha}$ are defined in (\ref{YL2}). 
To be able to bring the current in the FZ-multiplet form or the ${\cal R}$-multiplet form 
we have to make one more shift. 
First notice that 
\be
\bar  D^2 ( \bar Z \S )  =   \bar D^2 ( Y \S ) = Y  \bar D^2 ( \S ) =  Y   \bar D^2 \bar Y 
= (\frac1m \bar D^2 \bar \Phi) (m \Phi) = \bar D^2 (\Phi \bar \Phi)  
\ee
which gives
\be
\label{phiphi}
\bar  D^2 ( \bar Z \S ) = - \bar  D^2 ( \bar Z \S ) + 2 \bar D^2 (\Phi \bar \Phi)  . 
\ee
Now we insert (\ref{phiphi}) into (\ref{YYYYyy}) to find 
\be
{\cal Y}_{\alpha} = \frac12 D_\alpha\bar{D}^2 \ls  \Sigma\bar\Sigma - T - Z \bar \S - \bar Z \S  + \frac32 \Phi \bar \Phi  \rs. 
\ee
Now we are ready to perform appropriate shifts of type A with real $U$ to 
bring the current to the desired form.

To find the FZ-multiplet we perform a type A shift  with  
\be
U =  \frac16  \ls \Sigma\bar\Sigma - 3 T - Z \bar \S - \bar Z \S   + \frac12 \Phi \bar \Phi \rs
\ee
which gives a real ${\cal J}_{\alpha \dot \alpha}$ with 
\be
{\cal X}_\alpha &=& 0
\ee
and
\be
X = \frac13 \bar D^2 \ls  2 T  +  \Phi \bar \Phi  \rs . 
\ee 
It is clear that since we can bring the supercurrent conservation equation to  this 
form the model can be consistently coupled to the old-minimal supergravity.  
One may perform an appropriate shift and bring the system to the 
supercurrent conservation related to the new-minimal supergravity. 
To achieve this we perform a type A shift with 
\be
U = \frac12  \ls  \Sigma\bar\Sigma - T - Z \bar \S + \bar Z \S  - \frac12 \Phi \bar \Phi  \rs
\ee
which gives 
\be
{\cal X}_\alpha &=& \bar D^2 D_\alpha \ls  - \S \bar \S + Z \bar \S + \bar Z \S + \Phi \bar \Phi \rs 
\ee
and
\be
X = 0 . 
\ee

Now we can go to the IR limit for the FZ-multiplet. 
For the supersymmetry breaking vacuum, 
in the IR (and on-shell) as we explained
\be
\S^{(IR)} = X_{NL}
\ee
and
\be
Y^{(IR)}=0  \ , \  \Phi^{(IR)}=0 . 
\ee 
Then we have after a short calculation 
\be
X^{(IR)} =  \frac{1}{3} f  X_{NL} . 
\ee
We see again that $X$ in the low energy flows to $X_{NL}$ \cite{Komargodski:2009rz}.

\section{Goldstino description}

In this section  we 
focus on the supersymmetry breaking vacua, 
and give the low energy description of the  complex linear Goldstino superfield  in terms of the 
Samuel-Wess superfield $\L_\alpha$ \cite{Samuel:1982uh}. 
For the CNM multiplet (\ref{SMPHI}) with Lagrangian (\ref{L1}), 
we have shown that in the IR the massive sector will decouple, 
and leave  only the Goldstino sector behind. 
For the  complex linear, 
we have seen that one can employ mediation terms which will generically give non-supersymmetric masses to all the 
other modes except the Goldstino mode, 
therefore again in the IR there will be only the Goldstino. 
In other words for the complex linear model we employ both (\ref{L6}) and (\ref{MPSI}). 
Therefore, 
our models can be treated under a common framework in the IR, 
which is of course the concept of an effective low energy description; 
the UV properties of the theory are not important any more.

The $\L$-superfield \cite{Samuel:1982uh} satisfies the conditions 
\be
\label{sw1}
D_\beta \L_\alpha &=& \frac{1}{\kappa} C_{\alpha \beta} 
\\
\label{sw1'}
\bar D_{\dot \beta} \bar \L_{\dot \alpha} &=&  \frac{1}{\kappa} C_{\dot \alpha \dot \beta} 
\\
\label{sw2}
\bar D^{\dot \beta} \L^{\alpha} &=& i \kappa \L_\beta \p^{\beta \dot \beta} \L^\alpha 
\\
\label{sw2'}
D^{\beta} \bar \L^{\dot \alpha} &=& i \kappa \bar \L_{\dot \beta} \p^{\beta \dot \beta} \bar \L^{\dot \alpha} 
\ee
and $\kappa$ is related to the supersymmetry breaking scale 
($\kappa$ here is assumed to be real without loss of generality). 
The minimal superspace Lagrangian for the $\L$-superfield, 
has the form 
\be
\label{GOLD}
{\cal L}_{\L} = - \int d^4 \theta \, \L^\alpha \L_\alpha \bar \L^{\dot \alpha} \bar \L_{\dot \alpha} .
\ee
Before we turn to the complex linear Goldstino, 
let us review the chiral superfield Goldstino description. 
In this case the supersymmetry breaking Lagrangian is 
\be
\label{LLPHI}
{\cal L} = \int d^4 \theta \Phi \bar \Phi - \lc \int d^2 \theta  f \Phi  + c.c. \rc  
\ee
and the appropriate embedding of the Goldstino into the chiral superfield is 
\be
\label{LPHI} 
\Phi_\L = - \frac{\kappa}{2} \bar D^{\dot \beta} \bar D_{\dot \beta} ( \L^\alpha \L_\alpha \bar \L^{\dot \alpha} \bar \L_{\dot \alpha} ) 
\ee
for $f = -4 \kappa^{-3}$. 
If we insert (\ref{LPHI})  into (\ref{LLPHI}) we will find it is proportional to (\ref{GOLD}), 
with the correct sign.

It is also interesting to look at the modified complex linear superfield given in \cite{Kuzenko:2011ti}
which is dual to the chiral model given in (\ref{LLPHI}). The model is described by a superfield $\Gamma$ satisfying
the modified complex linear constraint
\be
\bar{D}^2 \Gamma = f\,.
\ee
In \cite{Kuzenko:2011ti} it was shown that for this model the Goldstino can be embedded into the
modified complex linear superfield as
\be\label{Sergei}
\Gamma  = -\frac{2}{f}\bar\Lambda^2
\ee
for $f^2 = \frac{1}{2\kappa^2}$. Inserting the ansatz (\ref{Sergei}) into the action gives the
kinetic term of the Goldstino with the correct sign. Since in this case we can write
\be
\bar\Lambda_{\dot\alpha} = -\frac{1}{\sqrt{2}}\bar{D}_{\dot\alpha} \Gamma
\ee
we see that it is the physical fermion that becomes the Goldstino as one may expect from the duality
with the chiral model. Because of the very simple relation between $\Gamma$ and
$\bar\Lambda_{\dot\alpha}$ in this model, one may invert equation (\ref{Sergei}) to express
the Samuel-Wess superfield in terms of $\Gamma$. Therefore it is possible to use $\Gamma$
as an alternative to $\Lambda_\alpha$ when one wants to describe superfield embeddings of the
Goldstino in any model\footnote{We would like to thank Sergei Kuzenko for discussions on this topic
\cite{Kuzenko:2015uca}.}.

As we have seen in our case,
at low energy, the only sector of the CNM and the complex linear models which does not decouple 
in the broken vacuum  is the Goldstino modes inside $\S$. 
We propose that the appropriate IR  description for $\S$ in the broken vacuum is 
\be
\label{SL}
\S_\L = \bar D^{\dot \alpha} \left( \bar \L_{\dot \alpha} \L^\alpha \L_\alpha  \right) 
\ee 
where $\L_\alpha$ is the Samuel-Wess Goldstino superfield \cite{Samuel:1982uh}.

Let us explain why (\ref{SL}) is the correct description in terms of $\L$. 
First we can see that 
\be
\bar D^2 \S_\L = 0 . 
\ee
Secondly, the Goldstino does not reside in the component $\bar D_{\dot \alpha} \S_\L |$ (the physical fermion), 
but rather in
\be
G_\alpha = D_\alpha \S_\L |   
\ee 
which is the  previously auxiliary fermion $\lambda_\alpha$. 
Moreover, 
we have 
\be
\langle F \rangle = \langle D^2 \S_\L | \rangle = -\frac{4}{\kappa^3}  
\ee
which also gives the relation to the supersymmetry breaking scale. 
In addition, notice that
\be
\label{S22}
\S_\L^2 = 0 . 
\ee
Finally, we can study the free Lagrangian for the complex linear superfield and replace $\S$ with the Goldstino 
superfield $\S_\L$. 
We have  
\be
\label{Lagrangian}
{\cal L} = - \int d^4 \theta \, \S_\L \bar \S_\L = 
- \frac{4}{\kappa^2} \int d^4 \theta \, \L^\alpha \L_\alpha \bar \L^{\dot \alpha} \bar \L_{\dot \alpha}  
\ee
with the right hand side being the standard Lagrangian for the Goldstino in the $\L$-superfield formulation \cite{Samuel:1982uh}. 
One might ask whether we could have $\S$ equal to $\bar D^2 (\L^2 \bar \L^2)$ in the IR. 
{}From the result in (\ref{Lagrangian}), 
one can understand that this would not be appropriate 
to describe a complex linear  Goldstino superfield, 
since this would lead to a Lagrangian for the $\L$-superfield 
with the wrong sign.

We now want to revisit the Lagrangian  
\be
\label{LSSSS}
{\cal L} = -\int d^4 \theta \ \bar \Sigma \Sigma + \frac{1}{8 f^2}  \int d^4 \theta  D^\alpha \Sigma D_\alpha \Sigma 
\bar D^{\dot{\beta}} \bar \Sigma  \bar D_{\dot{\beta}} \bar \Sigma  
\ee
for which we know there exists supersymmetry breaking vacua. 
As we explained, 
this Lagrangian is the low energy description in the broken vacuum for both 
the CNM and the complex linear models. 
For this model the Goldstino multiplet in the broken vacuum is described by $\S = \S_\L$ with
\be
f = -4 \kappa^{-3} . 
\ee 
Notice that the higher dimension operator becomes proportional to the standard kinetic term for $\S = \S_\L$ 
\be
 \frac{1}{8 f^2} \int d^4 \theta  D^\alpha \Sigma_\L D_\alpha \Sigma_\L 
 \bar D^{\dot{\beta}} \bar \Sigma_\L  \bar D_{\dot{\beta}} \bar \Sigma_\L 
 =  \frac{1}{2} \int d^4 \theta \, \S_\L \bar \S_\L  
\ee
similarly to what happens for the chiral model with a supersymmetry breaking superpotential.
The important point is that the final Lagrangian contains only the Goldstino and 
it has the correct (non-ghost) sign 
\be
{\cal L} = - \frac{1}{2} \int d^4 \theta \, \S_\L \bar \S_\L  . 
\ee
A simple calculation gives
\be
\langle V \rangle= \frac12 f^2 
\ee
therefore we find the same vacuum energy as for the models \eqref{L1} and \eqref{L66}.

Now we want to find the superspace equations of motion for the Goldstino superfield  $\Lambda_{\alpha}$. We may insert the complex linear Goldstino (\ref{SL})  in the equations of motion that arise from Lagrangian 
(\ref{LSSSS}). The equations for $\S_\L$ will be 
\be
\label{eqSL}
\S_\L = - \frac{\kappa^6}{64} \bar D^{\dot \alpha} 
\left( \bar D_{\dot \alpha} \bar \S_\L D^\alpha \S_\L D_\alpha \S_\L \right) . 
\ee
A manipulation of the right hand side of (\ref{eqSL}) using the properties 
of the $\L$-superfield reveals 
\be
\bar D^{\dot \alpha} 
\left( \bar D_{\dot \alpha} \bar \S_\L D^\alpha \S_\L D_\alpha \S_\L \right) 
= -  \frac{64}{\kappa^6} \Phi_\L 
\ee
which shows that the equations for $\S$ in fact predict that on-shell 
\be
\label{SeqX}
\S_\L =  \Phi_\L . 
\ee
Equation (\ref{SeqX}) is exactly the equation we had to assume such 
that we could solve the superspace equations of motion earlier (see for example (\ref{HeqXNL})). 
Of course  (\ref{eqSL}) is not satisfied by using only the $\L$-superfield properties, 
but it gives a restriction on $\L_\alpha$ 
\be
\label{Lambdaeq1}
\bar D^{\dot \alpha} \left( \L^2 \bar \L^2 \p^{\alpha}_{\dot \alpha} \L_{\alpha} \right) = 0 
\ee
which can be also written as
\be
\label{Lambdaeq2}
-\frac{4 i}{\kappa^2} \, \L^\beta \L_\beta \, \bar \L^{\dot \beta} \p_{\gamma \dot \beta} \L^{\gamma} 
+ \L^\beta \L_\beta \, \bar \L^{\dot \beta} 
\bar \L_{\dot \beta} \, \p^{\gamma \dot \rho} \p_{\gamma \dot \rho} \left( \L^\rho \L_\rho \right) =0
\ee
and represents the superspace equations of motion for the $\L$-superfield.  
The lowest component of the superspace equation (\ref{Lambdaeq2})  
can be shown to be compatible with the equations of the 
Goldstino fermion (the lowest component of $\L_\alpha$). 
Indeed, 
we may expand the Lagrangian (\ref{GOLD}) in components and perform a variation with respect to 
$\bar \L^{\dot \beta}| =\bar G^{\dot \beta}$.  
After we multiply with $G^2 \bar G^{\dot \beta}$ 
we have  
\be
-\frac{4 i}{\kappa^2} \, G^\beta G_\beta \, \bar G^{\dot \beta} \p_{\gamma \dot \beta} G^{\gamma} 
+ G^\beta G_\beta \, \bar G^{\dot \beta} 
\bar G_{\dot \beta} \, \p^{\gamma \dot \rho} \p_{\gamma \dot \rho} \left( G^\rho G_\rho \right) =0
\ee
and we compare with (\ref{Lambdaeq2}) to see that they are identical. 
This verifies that (\ref{Lambdaeq1}) is the $\L$-superfield equations of motion.

Finally, 
from equation \eqref{XNL2}, which as we said also gives equations of motion for the Goldstino, 
we get 
\be
\label{444}
\left( \bar \Phi_\L \bar D^2 \bar \Phi_\L -  f  \bar \Phi_\L \right) D_\alpha \Phi_\L =0  
\ee  
where we have replaced $X_{NL}$ with $\Phi_\L$ and multiplied with $\bar \Phi_\L  D_\alpha \Phi_\L$. 
Formula \eqref{444} is not trivially satisfied just from the 
properties of $\Phi_\L$, 
but rather it yields an additional equation for $\L$. 
Expanding \eqref{444}  in $\L$ gives 
\be
\L^2 \bar \L^2 \p^{\alpha}_{\dot \alpha} \L_{\alpha}  = 0 
\ee
which again implies \eqref{Lambdaeq1}.

\section{Conclusions}

In this work we have studied the properties 
of non-minimal multiplets as candidates for 
the hidden sector of supersymmetry breaking. 
We have explored the properties of two key models: 
the CNM multiplet and the complex linear multiplet with mediation terms. 
We have employed superspace higher derivatives, 
such that the auxiliary field potential is deformed and the system has acquires new 
supersymmetry breaking vacuum solutions. 
In these vacua, 
naively auxiliary fermionic fields become propagating 
and in particular they become the fermionic Goldstone modes. 
We have revisited the duality between non-minimal theories and chiral models 
and shown that the conventional duality procedure can not always capture the 
full dynamics of the theory, 
especially when auxiliary fields have become propagating - as happens here. 
Moreover, we have followed the Noether procedure for superdiffeomorphisms and 
we have identified the chiral $X$ superfield which enters the supercurrent  equations. 
For both models we have shown that in the IR it becomes the chiral Goldstino superfield $X_{NL}$. 
Finally, 
we have given a description for the Goldstino in terms of the Samuel-Wess $\L$-superfield, 
which works both  for the CNM and the complex linear model and therefore offers a universal description, 
and we have identified the superspace equations of motion.

\section*{Acknowledgments} 

We thank K. Bering, U. Lindstr\"om, M. Ro\v{c}ek and I. Sachs for discussions. 
This work is supported by the Grant agency of the Czech republic under the grant P201/12/G028.

\end{document}